\documentclass[12pt]{article}
\usepackage{amsmath}
\usepackage{amssymb}
\newfont{\elevenmib}{cmmib10 scaled\magstep1}%
\def\val{\alpha}
\def\al{\alpha}

\def\ga{\gamma}

\def\ep{\epsilon}

\def\la{\lambda}

\def\om{\omega}

\def\Ga{\Gamma}

\newcommand{\beq}{\begin{equation}}
\newcommand{\eeq}{\end{equation}}
\newcommand{\bea}{\begin{eqnarray*}}
\newcommand{\eea}{\end{eqnarray*}}
\newcommand{\beaq}{\begin{eqnarray}}
\newcommand{\eeaq}{\end{eqnarray}}
\makeatletter
\def\numberbysection{\@addtoreset{equation}{section}
\def\theequation{\thesection.\arabic{equation}}}
\makeatother

\numberbysection
\begin{document}
\centerline{\Large\bf ZZ-Branes of $N=2$ Super-Liouville Theory}
\vskip 2cm
\centerline{\large Changrim Ahn,\footnote{ahn@ewha.ac.kr}
Marian Stanishkov,\footnote{stanishkov@dante.ewha.ac.kr;
On leave of absence from Institute of Nuclear Research and Nuclear Energy,
Sofia, Bulgaria}
and Masayoshi Yamamoto\footnote{yamamoto@dante.ewha.ac.kr}}
\vskip 1cm
\centerline{\it Department of Physics}
\centerline{\it Ewha Womans University}
\centerline{\it Seoul 120-750, Korea}
\centerline{\small PACS: 11.25.Hf, 11.55.Ds}

\vskip 1cm
\centerline{\bf Abstract}
We study conformal boundary conditions and corresponding one-point functions
of the $N=2$ super-Liouville theory
using both conformal and modular bootstrap methods.
We have found both continuous (`FZZT-branes') and discrete (`ZZ-branes')
boundary conditions.
In particular, we identify two different types of the discrete ZZ-brane
solutions, which are associated with degenerate fields of the
$N=2$ super-Liouville theory.

\section{Introduction}

Two-dimensional Liouville field theory (LFT) has been studied
for its relevance with non-critical string theories and two-dimensional
quantum gravity \cite{LFT,CurTho}.
Recently, string theory with 2D Euclidean black hole geometry \cite{Witten}
has been claimed to be T-dual to the sine-Liouville theory \cite{FZZ2,KKK}.
A similar duality has been discovered for the $N=2$ supersymmetric
Liouville field theory (SLFT) and a fermionic 2D black hole which is
identified with super $SL(2,R)/U(1)$ coset
conformal field theory (CFT) \cite{GK,HoriKap}.

The LFT and SLFTs are irrational CFTs
which have continuously infinite number of primary fields.
The irrationality requires new formalisms for exact computation of
correlation functions.
One very efficient formalism is the conformal bootstrap
method which has been first applied to the LFT \cite{Teschner,ZamZam},
and later to the $N=1$ SLFT \cite{RasSta,Poghossian},
and the $N=2$ SLFT \cite{AKRS}.

Conformally invariant boundary conditions (BCs) for these models
have been also actively studied.
The conformal bootstrap method has been extended to the LFT with
boundary in \cite{FZZ,ZZ}, and the $N=1$ SLFT \cite{ARS,FH}.
Recently, the boundary conformal bootstrap of the $N=2$ SLFT has
been studied in \cite{ASY}.
While the LFT and $N=1$ SLFT are invariant under a dual transformation of
the coupling constant $b\to 1/b$,
the $N=2$ SLFT does not have this self-duality.
To complement this, a $N=2$ supersymmetric theory dual to the $N=2$ SLFT
was proposed along with various consistency checks in \cite{AKRS} and proved
in \cite{nakayama}.
This dual theory has provided additional functional equation which, along with
the equation based on the $N=2$ SLFT, can produce exact correlation functions
such as the reflection amplitudes.
However, this approach can not be extended to the boundary conformal
bootstrap method since the boundary action for
the dual $N=2$ theory is not easy to write down.
Due to this deficiency, the boundary conformal
bootstrap equations considered in \cite{ASY} contained an undetermined
coefficient and could be used only to check the consistency of
the one-point functions obtained by the `modular bootstrap' method.
Moreover, these equations were applicable to the `neutral' fields due to
technical difficulties.
One of our motivations in this paper is to derive more functional equations
based on the conformal bootstrap method and determine the one-point functions
exactly including those of `non-neutral' fields.

The modular bootstrap method is a generalization of the Cardy formulation
for the conformal BCs to the irrational CFTs.
This method has been initiated in \cite{ZZ} and extended to the $N=1$
SLFT \cite{ARS,FH}.
In these works, one-point function of a certain primary field under
a specific BC is associated with the boundary amplitude which is
the scalar product of the corresponding
Ishibashi state and the conformal BC state.
The boundary amplitudes satisfy the Cardy conditions which are expressed
in terms of the modular $S$-matrix elements.
While this method is proven to be effective, it has been mainly used
to check the consistency of the one-point functions derived by the
conformal bootstrap method because there are some ambiguities in
deriving the boundary amplitudes from the Cardy conditions.

For the $N=2$ SLFT, the situation becomes different.
Since the conformal bootstrap could not be completed,
the modular bootstrap method remains the only available formalism to derive
the boundary amplitudes for the $N=2$ SLFT and its T-dual,
$SL(2,R)/U(1)$ super-coset CFT in \cite{MMV,RibSch,EguSug,ASY,IKPT,ES2}.
Then, one-point functions of the `continuous' BC,
sometimes called `FZZT-brane', derived in this way
are confirmed by the conformal bootstrap equations \cite{ASY}.

To find all the consistent conformal BCs of the $N=2$ SLFT and its dual
super-coset model is important since they describe the D-branes moving in
the black hole background.
In addition to the FZZT and the vacuum BCs, there are infinite number
of discrete BCs, which are called `ZZ-branes' in general .
For the LFT \cite{ZZ} and $N=1$ SLFT \cite{ARS,FH},
the ZZ-branes have been constructed
in the background geometry of the classical Lobachevskiy plane
or the pseudosphere \cite{DJ}.
Our main result in this paper is to construct the general ZZ-brane BCs of
the $N=2$ SLFT.

In sect.2 we introduce the $N=2$ SLFT and its operator contents, in particular,
degenerate fields and their properties.
Based on this, we derive the functional equations for the one-point functions
of primary fields for the continuous BC (FZZT) in sect.3.
In sect.4, we find the ZZ-brane solutions from the functional equations defined on
the pseudosphere and discuss their implications for the $N=2$ SLFT.
Modular bootstrap computations for the degenerate fields of the $N=2$ SLFT
are performed in sect.5 which provides some consistency checks for the
solutions.
We conclude this paper with some remarks in sect.6.

\section{$N=2$ Super-Liouville Theory}

In this section, we introduce the action and the $N=2$ superconformal algebra
and the primary fields.
In particular we discuss the degenerate fields in detail.

\subsection{$N=2$ superconformal algebra}

The action of the $N=2$ SLFT is given by
\beaq
S&=&\int d^2z\Bigg[\frac{1}{2\pi}\left(\partial\phi^-\bar{\partial}\phi^+
+\partial\phi^+\bar{\partial}\phi^-
+\psi^-\bar{\partial}\psi^++\psi^+\bar{\partial}\psi^-
+\bar{\psi}^-\partial\bar{\psi}^++\bar{\psi}^+\partial\bar{\psi}^-\right)
\nonumber\\
&&+i\mu b^2\psi^-\bar{\psi}^-e^{b\phi^+}
+i\mu b^2\psi^+\bar{\psi}^+e^{b\phi^-}
+\pi\mu^2 b^2e^{b(\phi^++\phi^-)}\Bigg].
\label{N2L}
\eeaq
This theory needs a background charge $1/b$ for conformal invariance so that
the interaction terms in Eq.(\ref{N2L}) become the screening
operators of the CFT.

The stress tensor $T$, the supercurrents $G^{\pm}$ and
the $U(1)$ current $J$ of this CFT are given by
\beaq
T&=&-\partial\phi^-\partial\phi^+
-\frac{1}{2}(\psi^-\partial\psi^++\psi^+\partial\psi^-)
+{1\over{2b}}(\partial^2\phi^++\partial^2\phi^-),
\label{N2T}\\
G^{\pm}&=&\sqrt{2}i(\psi^{\pm}\partial\phi^{\pm}-{1\over{b}}
\partial\psi^{\pm}),\qquad
J=-\psi^-\psi^{+}+{1\over{b}}(\partial\phi^+-\partial\phi^-).
\label{N2J}
\eeaq
The $N=2$ super-Virasoro algebra is expressed by the modes of these currents
as follows:
\bea
\left[L_m,L_{n}\right]&=&(m-n)L_{m+n}+{c\over{12}}(m^3-m)\delta_{m+n},\\
\left[L_m,G^{\pm}_{r}\right]&=&\left({m\over{2}}-r\right)G^{\pm}_{m+r},
\qquad \left[J_n,G^{\pm}_{r}\right]=\pm G^{\pm}_{n+r},\\
\left\{G^{+}_{r},G^{-}_{s}\right\}&=&2L_{r+s}+(r-s)J_{r+s}
+{c\over{3}}\left(r^2-{1\over{4}}\right)\delta_{r+s},\quad
\left\{G^{\pm}_{r},G^{\pm}_{s}\right\}=0,\\
\left[L_m,J_{n}\right]&=&-nJ_{m+n},\qquad
\left[J_m,J_{n}\right]={c\over{3}}m\delta_{m+n},
\eea
with the central charge
\beq
c=3+6/b^2.
\eeq

Super-CFTs have both the Neveu-Schwarz (NS) sector with half-integer
fermionic modes and the Ramond (R) sector with integer modes.
The primary fields of the $N=2$ SLFT are also classified into the
(NS) and the (R) sectors and can be written as follows \cite{Marian}:
\begin{equation}
N_{\alpha{\overline\alpha}}=
e^{\alpha\phi^{+}+{\overline\alpha}\phi^{-}},\qquad
R^{(\pm)}_{\alpha{\overline\alpha}}=
\sigma^{\pm}e^{\alpha\phi^{+}+{\overline\alpha}\phi^{-}},
\label{primary}
\end{equation}
where $\sigma^{\pm}$ are the spin operators.

The conformal dimensions and the $U(1)$ charges of the primary fields
$N_{\alpha{\overline\alpha}}$ and $R^{(\pm)}_{\alpha{\overline\alpha}}$
can be obtained:
\begin{equation}
\Delta^{NS}_{\alpha{\overline\alpha}}=-\alpha{\overline\alpha}
+{1\over{2b}}(\alpha+{\overline\alpha}),\qquad
\Delta^{R}_{\alpha{\overline\alpha}}=\Delta^{NS}_{\alpha{\overline\alpha}}
+{1\over{8}},
\label{delta}
\end{equation}
and
\begin{equation}
\omega={1\over{b}}(\alpha-{\overline\alpha}),
\qquad
\omega^{\pm}=\omega \pm{1\over{2}}.
\label{u1charge}
\end{equation}
It is more convenient sometimes to use a `momentum' defined by
\beq
\alpha+{\overline\alpha}={1\over{b}}+2iP, \label{momentum}
\eeq
and the $U(1)$ charge $\omega$ instead of $\alpha,{\overline\alpha}$.
In terms of these, the conformal dimension is given by
\begin{equation}
\Delta^{NS}={1\over{4b^2}}+P^2+{b^2\omega^2\over{4}}.
\label{deltanew}
\end{equation}
We will denote the primary fields by $N_{[P,\omega]}$ and
$R^{(\pm)}_{[P,\omega]}$ in sect.5.

\subsection{Degenerate fields}

Among the primary fields there is a series of degenerate fields of the $N=2$ SLFT.
In this paper we divide these fields into three classes.
Class-I degenerate fields are given by
\begin{eqnarray}
N_{m,n}^{\omega}&=&N_{\alpha_{m,n}^{\omega},{\overline\alpha}_{m,n}^{\omega}},
\qquad
R_{m,n}^{(\ep)\omega}=R^{(\ep)}_{\alpha_{m,n}^{\omega},
{\overline\alpha}_{m,n}^{\omega}},\\
\alpha_{m,n}^{\omega}&=&{1-m+\omega b^2\over{2b}}-{nb\over{2}},\qquad
{\overline\alpha}_{m,n}^{\omega}={1-m-\omega b^2\over{2b}}-{nb\over{2}},
\ \ m,n\in {\mathbf Z}_{+}.
\label{degtypeI}
\end{eqnarray}
$N_{m,n}^{\omega}$ and $R_{m,n}^{(\ep)\omega}$ are degenerate at the level $mn$
where the corresponding null states are turned out to be
\beq
N_{m,-n}^{\omega},\qquad{\rm and}\qquad R_{m,-n}^{(\ep)\omega}.
\label{nullone}
\eeq
As an example, consider the most simple case $N_{1,1}^{\omega}$ with
the conformal dimension $b^2(\omega^2-1)/4-1/2$ and $U(1)$ charge $\omega$.
After simple calculation, one can check that
\beq
\left[{b^2\over{2}}(1-\omega^2)J_{-1}+G^{+}_{-1/2}G^{-}_{-1/2}
-(1-\omega)L_{-1}\right]\vert N_{1,1}^{\omega}\rangle
\eeq
is annihilated by all the positive modes of the $N=2$ super CFT.
Since this state has the $U(1)$ charge $\omega$ and dimension $+1$ more than
that of $N_{1,1}^{\omega}$, it corresponds to
$\vert N_{1,-1}^{\omega}\rangle$ up to a normalization constant.
One can continue this analysis to higher values of $m,n>1$ to confirm the
statement of Eq.(\ref{nullone}).
Notice that the null state structure changes dramatically for $\omega=\pm n$
case.
The field $N_{m,n}^{\pm n}$ has a null state $N_{m,-n}^{\pm n}$ at level $mn$.
This $N_{m,-n}^{\pm n}$ field is in fact a class-II degenerate field
which we will explain next and has infinite number of null states.
Therefore, we exclude the case of $\omega=\pm n$ from class-I fields.

The second class of degenerate fields is denoted by
$N_{m}^{\omega}$ and $R_{m}^{(\ep)\omega}$ and comes in two subclasses,
namely, class-IIA and class-IIB.
These are given by
\beaq
{\rm Class-IIA}&:&\qquad N_{m}^{\omega}=N_{\alpha_{m}^{\omega},
{\overline\alpha}_{m}^{0}}\qquad
R_{m}^{(+)\omega}=R^{(+)}_{\alpha_{m}^{\omega},{\overline\alpha}_{m}^{0}},
\quad \om>0
\label{degtypeIIa}\\
{\rm Class-IIB}&:&\qquad {\tilde N}_{m}^{\omega}=
N_{\alpha_{m}^{0},{\overline\alpha}_{m}^{\omega}}\qquad
R_{m}^{(-)\omega}=R^{(-)}_{\alpha_{m}^{0},{\overline\alpha}_{m}^{\omega}},
\quad \om<0.
\label{degtypeIIb}
\eeaq
Here we have defined
\beq
\alpha_{m}^{\omega}\equiv{1-m+2\omega b^2\over{2b}},\qquad
{\overline\alpha}_{m}^{\omega}\equiv{1-m-2\omega b^2\over{2b}}
\eeq
with $m$ a positive odd integer for the (NS) sector and
even for the (R) sector.

These fields have null states at level $m/2$ which can be
expressed again by Eq.(\ref{degtypeIIa}) with
$\omega$ shifted by $+1$ for class-IIA
and by Eq.(\ref{degtypeIIb}) with $\omega$ shifted by $-1$ for class-IIB.
For $m=1$, these fields become either chiral or
anti-chiral field which are annihilated by $G^{\pm}_{-1/2}$,
respectively.
For $m=3$, one can construct a linear combination of descendants
\beq
\left[\left(\omega-{2\over{b^2}}+1\right)G^{+}_{-3/2}
-G^{+}_{-1/2}L_{-1}+G^{+}_{-1/2}J_{-1}\right]\vert
N_{3}^{\omega}\rangle
\label{nullex}
\eeq
which satisfies the null state condition.
Since this state has $U(1)$ charge $\omega+1$
and dimension $3/2$ higher than that of $N_{3}^{\omega}$, it is
straightforward to identify it as $N_{3}^{\omega+1}$ up to a
normalization constant.
However, it is not the end of the story in this case.
The $N_{3}^{\omega+1}$ field is again degenerate at level $3/2$
because a linear combination of its descendants, exactly
Eq.(\ref{nullex}) with $\omega$ shifted by $+1$, satisfies the
null state condition.
This generates $N_{3}^{\omega+2}$ and it continues infinitely.
This infinite null state structure holds for any odd integer $m$.

This can be illustrated by semi-infinite sequences,
\beaq
{\rm Class-IIA}&:& N_{m}^{\omega}\to N_{m}^{\omega+1}\to
N_{m}^{\omega+2}\to\ldots\\
{\rm Class-IIB}&:&{\tilde N}_m^{\omega}\to {\tilde N}_m^{\omega-1}\to
{\tilde N}_m^{\omega-2}\to\ldots.
\eeaq
This works similarly for the (R) sector.
For example, the null state of the $m=2$ (R) field is given by
\beq
G^{\pm}_{-1}\vert R_{2}^{(\pm)\omega}\rangle.
\eeq

We need to deal with class-II neutral ($\omega=0$) (NS) fields separately.
For example, consider the $N_{3}^{0}$ which has two null states
\beq
\left[\left(1-{2\over{b^2}}\right)G^{\pm}_{-3/2}
-G^{\pm}_{-1/2}L_{-1}+G^{\pm}_{-1/2}J_{-1}\right]\vert N_{3}^{0}\rangle,
\label{nullexi}
\eeq
which should be identified with $N_{3}^{1}$ and ${\tilde N}_{3}^{-1}$,
respectively.
We will call these neutral (NS) degenerate fields as class-III and denote by
\beq
{\rm Class-III}:\qquad N_{m}=N_{\alpha_{m}^{0}{\overline\alpha}_{m}^{0}}.
\label{degtypeIII}
\eeq
The null state structure of the class-III fields has an infinite sequence
in both directions,
\beq
{\rm Class-III}:\qquad \ldots\leftarrow{\tilde N}_m^{-2}\leftarrow
{\tilde N}_m^{-1}\leftarrow N_{m}\to N_{m}^{1}\to N_{m}^{2}\to\ldots.
\label{neutralseq}
\eeq
The identity operator is the most simple class-III field with $m=1$.

The degenerate fields are playing an essential role in
both conformal and modular bootstraps.
As we will see shortly, some simple degenerate fields satisfy
relatively simple operator product expansion (OPE) and make the conformal
bootstrap viable.
In this paper we will associate the conformal BCs corresponding to
the degenerate fields with solutions of the functional equations obtained
by the conformal bootstrap.

\section{FZZT-Branes}

The FZZT-branes can be described as the $N=2$ SLFT on
a half-plane whose BCs are characterized by a continuous parameter.
Extending our previous work \cite{ASY},
we complete the conformal bootstrap in this section.

\subsection{One-point functions}

In this section, we compute exact one-point functions of the (NS) and (R)
bulk operators $N_{\alpha{\overline\alpha}}$ and
$R^{(+)}_{\alpha{\overline\alpha}}$ of the
SLFT with boundary.
\footnote{From now on we consider $\ep=+$ only since the other
case is almost the same.}
The boundary preserves the $N=2$ superconformal symmetry
if the following boundary action is added \cite{AY}:
\beaq
S_B&=&\int_{-\infty}^{\infty}dx\Bigg[-\frac{i}{4\pi}
(\bar{\psi}^+\psi^-+\bar{\psi}^-\psi^+) +\frac{1}{2}a^-\partial_x a^+
\nonumber\\
&-&\frac{1}{2}e^{b\phi^+/2}\left(\mu_B a^+
+\frac{\mu b^2}{4\mu_B}a^-\right)(\psi^-+\bar{\psi}^-)
-\frac{1}{2}e^{b\phi^-/2}\left(\mu_B a^-
+\frac{\mu b^2}{4\mu_B} a^+\right)(\psi^++\bar{\psi}^+)\nonumber\\
&-&\frac{2}{b^2}\left(\mu_B^2+\frac{\mu^2b^4}{16\mu_B^2}\right)
e^{b(\phi^++\phi^-)/2}\Bigg].
\label{baction}
\eeaq

The one-point functions are defined by
\beq
\langle N_{\alpha{\overline\alpha}}(\xi,\bar\xi)\rangle
={U^{NS}(\alpha,{\overline\alpha})\over
{|\xi-\bar\xi|^{2\Delta^{NS}_{\alpha{\overline\alpha}}}}},\quad
{\rm and}\quad
\langle R^{(+)}_{\alpha{\overline\alpha}}(\xi,\bar\xi)\rangle
={U^{R}(\alpha,{\overline\alpha})\over
{|\xi-\bar\xi|^{2\Delta^R_{\alpha{\overline\alpha}}}}},
\eeq
with the conformal dimensions given in Eq.(\ref{delta}).
We will simply refer to the coefficients $U^{NS}(\alpha,{\overline\alpha})$
and $U^R(\alpha,{\overline\alpha})$ as the one-point functions.

\subsection{Conformal bootstrap of the $N=2$ SLFT}

The conformal bootstrap method starts with a two-point correlation
function on the half-plane.
By choosing one of the two operators as a simple degenerate field,
the OPE relation becomes relatively simple.
The bootstrap equations arise from considering two different channels; one is
taking the OPE before the fields approach on the boundary and
the other channel is taking the degenerate field on the boundary where the
boundary screening integral based on the boundary action is considered
\cite{FZZ}.

In practice, due to technical difficulties, we could consider only a few
most simple degenerate fields and their OPEs.
Usually, these are enough to fix the one-point functions exactly up to
overall constants.
Let us first consider a two-point function of
a neutral degenerate field $N_{-b/2}$ and a general neutral field $N_{\alpha}$:
\footnote{
We will suppress one of the indices of the fields since ${\overline\alpha}=\alpha$.}
\beq
G_{\al}(\xi,\xi')=\langle N_{-b/2}(\xi)N_\val(\xi')\rangle.
\label{twopt}
\eeq
The product of these two fields are expanded into four fields
\beq
N_{-b/2} N_\val= \left[N_{\val-{b\over 2}}\right] + C_{+-}
\left[\psi^{+}{\bar\psi}^{+} N_{\val-{b\over 2},\val+{b\over 2}}\right]
+C_{-+}\left[\psi^{-}{\bar\psi}^{-} N_{\val+{b\over 2},\val-{b\over 2}}\right]
+C_{--}\left[N_{\val+{b\over 2}}\right].
\label{ope}
\eeq
Here the bracket [\ldots] means the conformal tower of a given primary field.
One can see that the second and third terms in the RHS are the
super-partners of the corresponding fields.
The structure constants can be computed by screening integrals
as follows:
\bea
C_{+-}(\al)&=&C_{-+}(\al)=-\pi\mu{\ga(\al b-{b^2\over 2}-1)\over{
\ga(-{b^2\over 2})\ga(\al b)}}\\
C_{--}(\alpha)&=&2^{-2b^2-2}\pi^2\mu^2 b^4
\ga(1-\val b)\ga\left({1\over 2}-{b^2\over 2}-\val b\right)
\ga\left(\val b-{1\over 2}\right)
\ga\left(\val b+{b^2\over 2}\right),
\eea
with $\ga(x)\equiv\Ga(x)/\Ga(1-x)$.

Using these, we can express the two-point function as
\beq
G_{\al}(\xi,\xi')=U^{NS}\left(\alpha-{b\over{2}}\right)
{\cal G}^{NS}_{1}(\xi,\xi')+C_{--}(\alpha)U^{NS}
\left(\alpha+{b\over{2}}\right) {\cal G}^{NS}_{3}(\xi,\xi')
\eeq
where the one-point functions of the super-partners
vanish due to the supersymmetric boundary.
The ${\cal G}^{NS}_{i}(\xi,\xi')$'s are expressed
in terms of the special conformal blocks
\beq
{\cal G}^{NS}_{i}(\xi,\xi')={|\xi'-{\bar \xi'}|^{2\Delta^{NS}_{\alpha}
-2\Delta^{NS}_{-b/2}}\over{|\xi-{\bar \xi'}|^{4\Delta^{NS}_{\alpha}}}}
{\cal F}^{NS}_{i}(\eta),\quad i=1,2,3
\nonumber
\eeq
with
\beq
\eta={(\xi-\xi')({\bar \xi}-{\bar \xi'})\over{(\xi-{\bar \xi'})
({\bar \xi}-\xi')}}.
\nonumber
\eeq
These conformal blocks can be determined by Dotsenko-Fateev integrals
\cite{DotFat}:
\beaq
I_{i}(\eta)&=&N_{i}{\cal F}^{NS}_{i}(\eta),\nonumber\\
&=&\int\int_{{\cal C}_i} dx_1dx_2\langle N_{-b/2}(\eta)N_\val(0)N_{-{b/2}}(1)
\psi^{-}e^{b\phi^{+}(x_1)}\psi^{+}e^{b\phi^{-}(x_2)}N_{1/b-\val}(\infty)\rangle
\nonumber\\
&=&\eta^{\val b}(1-\eta)^{-b^2/2}\int\int dx_1 dx_2 (x_1x_2)^{-\val b}\nonumber\\
&\times&[(x_1-1)(x_2-1)(x_1-\eta)(x_2-\eta)]^{b^2/2}(x_1-x_2)^{-b^2-1}.
\label{Iintdef}
\eeaq
The index $i$ denotes the three independent integration contours between
the branching points $0,\eta,1,\infty$.
The conformal blocks ${\cal F}^{NS}_i(\eta)$ are regular at $\eta=0$.
Since we are interested in the limit $\eta\rightarrow 1$,
we need to introduce another blocks which are well defined in that limit.
This can be provided by ${\tilde I}_{i}$ which is given by the same integral
(\ref{Iintdef}) with a different contour as explained in \cite{DotFat}.
Using this, we can define another set of conformal blocks as
\beq
\quad {\tilde I}_{i}(\eta)={\tilde N}_{i}{\tilde{\cal F}}^{NS}_{i}(\eta)
\eeq
so that ${\tilde {\cal F}}^{NS}_j(\eta)$ are regular at $\eta=1$.
The monodromy relations between the conformal blocks are given by \cite{DotFat}:
\beq
{\cal F}^{NS}_i(\eta)=\sum_{j=1}^{3}\val_{ij}{\tilde {\cal F}}^{NS}_j(\eta),
\eeq
with
\beaq
\val_{13}&=& {\Gamma(-1)\Gamma(-{1\over 2}-{b^2\over
2})\over \Gamma(-{b^2\over 2})\Gamma({1\over
2})}{\Gamma(\val b-{b^2\over 2}) \Gamma(\val b+{1\over 2})\over
\Gamma(\val b)
\Gamma(\val b-{b^2\over 2}-{1\over 2})}\nonumber\\
\val_{23}&=& {2\Ga(-1)\Gamma(-1-b^2)\over \Ga(-{b^2\over 2})
\Ga(1+{b^2\over 2})} {\Ga(\val b+{b^2\over 2}+1)\Ga(2-\val b+{b^2\over 2})
\over \Ga(\val b)\Ga(-\val b+1)}\\
\val_{33}&=& {\Ga(-1)\Ga(-{1\over 2}-{b^2\over 2})\over \Ga(-{b^2\over 2})
\Ga({1\over 2})}{\Ga({3\over 2}-\val b)\Ga(1-{b^2\over 2}-\val b)\over
\Ga(1-\val b)\Ga({1\over 2}-{b^2\over 2}-\val b)}.
\nonumber
\eeaq
Here we have written only those for ${\tilde{\cal F}}^{NS}_3$ because
we are interested in the identity operator in the intermediate channel.
Notice that this calculation involves a divergent constant $\Ga(-1)$.
We will show that this factor is canceled in the functional equation.

The two-point function $G_{\al}(\xi,\xi')$ in the other channel can be computed
as the two fields approach the boundary.
When the degenerate field $N_{-b/2}$ approaches, it can be expanded in
boundary operators including the boundary identity operator.
For the boundary identity operator, a special bulk-boundary
structure constant ${\cal R}(-b/2)$ can be computed by the boundary
action (\ref{baction}):
\beaq
{\cal R}(-b/2)&=&-{\overline\mu}_{B}^2\int\int dx_1dx_2
\langle N_{-{b\over{2}}}(i/2)\psi^{-}e^{b\phi^{+}(x_1)/2}
\psi^{+}e^{b\phi^{-}(x_2)/2}e^{{1\over{2b}}(\phi^{+}+\phi^{-})(\infty)}
\rangle\nonumber\\
&=&-2^{-b^2+1}\sqrt{\pi}{\overline\mu}_{B}^2
{\Ga(0)\Ga(-{1\over 2}-{b^2\over 2})\over{\Ga(-{b^2\over 2})}}
\eeaq
with
\beq
{\overline\mu_B}^2=\mu_B^2+\frac{\mu^2b^4}{16\mu_B^2}.
\label{bcosmo}
\eeq
This constant also contains a singular factor $\Ga(0)$.
After dividing out these factors on the both sides of the equation,
we obtain the following bootstrap equation
\beaq
2^{-b^2+1}\pi{\overline\mu}_{B}^2U^{NS}(\val)
&=&{\Ga(\val b-{b^2\over 2})\Ga(\val b+{1\over 2})\over{
\Ga(\val b)\Ga(\val b-{b^2\over 2}-{1\over 2})}}U^{NS}
\left(\val-{b\over 2}\right)\nonumber\\
&+&2^{-2-2b^2}\pi^2 b^4\mu^2 {\Ga(\val b-{1\over 2})\Ga(\val b+{b^2\over 2})
\over{\Ga(\val b)\Ga(\val b+{b^2\over 2}+{1\over 2})}}
U^{NS}\left(\val+{b\over 2}\right).
\label{funeqNi}
\eeaq
It turns out that a similar functional equation for the (R) field can not be
obtained in this way.
Instead we will show shortly that the other functional equations can be used to
find (R) one-point functions.

\subsection{Conformal bootstrap based on the dual action}

In \cite{AKRS} it was proposed that the $N=2$ SLFT is dual to the theory
with the action
\beaq
S&=&\int d^2z\Bigg[\frac{1}{2\pi}(\partial\phi^-\bar{\partial}\phi^++
\partial\phi^+\bar{\partial}\phi^-
+\psi^-\bar{\partial}\psi^++\psi^+\bar{\partial}\psi^-
+\bar{\psi}^-\partial\bar{\psi}^++\bar{\psi}^+\partial\bar{\psi}^-)
\nonumber\\
&&+\frac{\tilde{\mu}}{b^2}(\partial\phi^--\frac{1}{b}\psi^-\psi^+)
(\bar{\partial}\phi^+-\frac{1}{b}\bar{\psi}^+\bar{\psi}^-)
e^{\frac{1}{b}(\phi^++\phi^-)}\Bigg],
\label{dual}
\eeaq
where $\tilde{\mu}$ is the dual cosmological constant.
One can derive functional equations for the one-point functions of
neutral fields by using the screening operator in the dual action.

We consider the two-point function of the class-II degenerate field
$R^+_{-1/2b}$ and a (NS) primary field $N_{\alpha}$
\beq
G^{NS}_{\alpha}(\xi,\xi')=\langle R^+_{-\frac{1}{2b}}(\xi)N_{\alpha}(\xi')\rangle.
\label{GNS}
\eeq
The OPE of these fields is given by
\beq
R^+_{-\frac{1}{2b}}N_{\alpha}=\left[R^+_{\alpha-\frac{1}{2b}}\right]
+C^{NS}(\alpha)\left[R^+_{\alpha+\frac{1}{2b}}\right],
\label{RNOPE}
\eeq
where the structure constant $C^{NS}(\alpha)$ was computed in \cite{AKRS}
based on the dual action (\ref{dual})
\beq
C^{NS}(\alpha)=\pi\tilde{\mu}\gamma\left(1+b^{-2}\right)
\frac{\Gamma(\frac{2\alpha}{b}-\frac{1}{b^2})\Gamma(1-\frac{2\alpha}{b})}
{\Gamma(1-\frac{2\alpha}{b}+\frac{1}{b^2})\Gamma(\frac{2\alpha}{b})}.
\label{CNS}
\eeq
The two-point function (\ref{GNS}) can be written as
\beq
G^{NS}_{\alpha}(\xi,\xi')=\frac{|\xi'-\bar{\xi}'|^{2\Delta^{NS}
_{\alpha}-2\Delta^R_{-1/2b}}}{|\xi-\bar{\xi}'|^{4\Delta^{NS}_{\alpha}}}
\Bigg[U^R\left(\alpha-\frac{1}{2b}\right){\cal F}^{NS}_+(\eta)
+C^{NS}(\alpha)U^R\left(\alpha+\frac{1}{2b}\right){\cal F}^{NS}_-(\eta)\Bigg],
\label{GNS1}
\eeq
where ${\cal F}^{NS}_{\pm}(\eta)$ are special conformal blocks.
These conformal blocks can be obtained by the following integral
\beaq
&&\int dx\langle R^+_{-\frac{1}{2b}}(\eta)N_{\alpha}(0)R^-_{-\frac{1}{2b}}(1)
\psi^-(x)\psi^+(x)N_{\frac{1}{b}}(x)N_{\frac{1}{b}-\alpha}(\infty)\rangle
\nonumber\\
&&=\eta^{\frac{\alpha}{b}}(1-\eta)^{-\frac{1}{2b^2}+\frac{3}{4}}
\int dxx^{-\frac{2\alpha}{b}}(x-1)^{\frac{1}{b^2}-1}(x-\eta)^{\frac{1}{b^2}-1}.
\label{integral}
\eeaq
Due to two independent contours, this is expressed in terms of the
hypergeometric functions
\beaq
{\cal F}^{NS}_+(\eta)&=&\eta^{\frac{\alpha}{b}}(1-\eta)^{-\frac{1}{2b^2}
+\frac{3}{4}}F\left(-\frac{1}{b^2}+1,\frac{2\alpha}{b}-\frac{2}{b^2}+1;
\frac{2\alpha}{b}-\frac{1}{b^2}+1;\eta\right),
\label{FNS+}\\
{\cal F}^{NS}_-(\eta)&=&\eta^{-\frac{\alpha}{b}+\frac{1}{b^2}}
(1-\eta)^{-\frac{1}{2b^2}+\frac{3}{4}} F\left(-\frac{1}{b^2}+1,
-\frac{2\alpha}{b}+1;-\frac{2\alpha}{b}+\frac{1}{b^2}+1;\eta\right).
\label{FNS-}
\eeaq
In the cross channel, the two-point function can be written as
\beq
G^{NS}_{\alpha}(\xi,\xi')=\frac{|\xi'-\bar{\xi}'|^{2\Delta^{NS}_{\alpha}
-2\Delta^R_{-1/2b}}}{|\xi-\bar{\xi}'|^{4\Delta^{NS}_{\alpha}}}
[B^{NS}_+(\alpha)\tilde{\cal F}^{NS}_+(\eta)
+B^{NS}_-(\alpha)\tilde{\cal F}^{NS}_-(\eta)],
\label{GNS2}
\eeq
where $B^{NS}_{\pm}(\alpha)$ are the bulk-boundary structure constants
and $\tilde{\cal F}^{NS}_\pm(\eta)$ are given by
\beaq
\tilde{\cal F}^{NS}_+(\eta)&=&\eta^{\frac{\alpha}{b}}(1-\eta)^{-\frac{1}{2b^2}
+\frac{3}{4}}F\left(-\frac{1}{b^2}+1,\frac{2\alpha}{b}-\frac{2}{b^2}+1;
-\frac{2}{b^2}+2;1-\eta\right),
\label{tilFNS+}\\
\tilde{\cal F}^{NS}_-(\eta)&=&\eta^{\frac{\alpha}{b}}(1-\eta)^{\frac{3}{2b^2}
-\frac{1}{4}}F\left(\frac{2\alpha}{b},\frac{1}{b^2};\frac{2}{b^2};1-\eta\right).
\label{tilFNS-}
\eeaq
The conformal block $\tilde{\cal F}^{NS}_-(\eta)$ corresponds to the boundary
identity operator which appears in the boundary as the bulk operator
$R^+_{-1/2b}$ approaches the boundary.
The fusion of $N_{\alpha}$ to the boundary identity operator is described
by the one-point function $U^{NS}(\alpha)$.
On the other hand, the fusion of $R^+_{-1/2b}$ is described by
a special bulk-boundary structure constant ${\cal R}(-1/2b)$ which could be
computed as a boundary screening integral
with one insertion of the dual boundary interaction if it were known.
Therefore, the bulk-boundary structure constant $B^{NS}_-(\alpha)$ can be written
as $B^{NS}_-(\alpha)={\cal R}(-1/2b)U^{NS}(\alpha)$.

Comparing (\ref{GNS1}) with (\ref{GNS2}) and using the monodromy relations
between ${\cal F}_{\pm}(\eta)$ and $\tilde{\cal F}_{\pm}(\eta)$,
we obtain the following functional equation for the one-point function
\beaq
{\cal R}\left(-1/2b\right)U^{NS}(\alpha)
&=&\frac{\Gamma(\frac{2\alpha}{b}-\frac{1}{b^2}+1)\Gamma(-\frac{2}{b^2}+1)}
{\Gamma(-\frac{1}{b^2}+1)\Gamma(\frac{2\alpha}{b}-\frac{2}{b^2}+1)}
U^R\left(\alpha-\frac{1}{2b}\right)
\nonumber\\
&+&\pi\tilde{\mu}\gamma\left(1+b^{-2}\right)
\frac{\Gamma(\frac{2\alpha}{b}-\frac{1}{b^2})\Gamma(-\frac{2}{b^2}+1)}
{\Gamma(-\frac{1}{b^2}+1)\Gamma(\frac{2\alpha}{b})}
U^R\left(\alpha+\frac{1}{2b}\right).
\label{funceqdual}
\eeaq
In a similar way, one can derive a functional equation which relates $U^{R}(\alpha)$
with $U^{NS}(\alpha\pm 1/2b)$ as is given in \cite{ASY}.

\subsection{Conformal bootstrap based on the $N=2$ minimal CFTs}

So far, we have considered only the neutral one-point functions.
To consider the charge sector, we need to consider a non-neutral
degenerate field.
The most simple degenerate field is $N_{-1/b,0}$ which is a special
case of class-II field (\ref{degtypeIIb}) with $m=3$ which has a null state
at level $3/2$.
To use the usual conformal bootstrap screening procedure,
we need to consider the dual action proposed in \cite{AKRS} as in previous
subsection.
However, calculations will be quite complicated.
So, we adopt here another strategy which is based on analytic continuation
of the $N=2$ super-minimal CFTs investigated in \cite{Marian}.
These CFTs have central charges $c=3-{6\over {p+2}}$ and their
(NS) primary fields are denoted by two integers $l=0,1,...$ and
$m=-l, -l+2,...,l$.
Comparing the central charge of the $N=2$ SLFT, we can identify $b^2=-(p+2)$.
Furthermore, we can relate the primary fields of the two theories
by comparing the conformal dimensions and $U(1)$ charges as follows:
\beq
N^l_m=e^{-{{l+m}\over 2b}\phi^+ - {{l-m}\over 2b}\phi^{-}}
\label{Nlm}
\eeq
with
\beq
l=-b(\val +\bar\val),\qquad m=b(\bar\val-\val).
\label{l-m}
\eeq

The degenerate chiral field $N_{-1/b,0}$ is identified with $N^1_1$.
Its OPE with an arbitrary (NS) field is given by
\beq
N^1_1 N^l_m= N^{l+1}_{m+1} + C_-^{(NS)}N^{l-1}_{m+1}
\eeq
and can be translated into that of the $N=2$ SLFT
\beq
N_{-1/b,0}N_{\val,\bar\val}=N_{\val-1/b,\bar\val}
+ C_{-}^{(NS)}N_{\val,\bar\val +1/b}.
\label{ope2}
\eeq

Indeed, one can check that the structure constants in both cases coincide
if Eq.(\ref{l-m}) is imposed.
To write a functional equation for $U(\val,\bar\val)$,
we follow the same procedure as before.
The conformal blocks corresponding to the two terms in Eq.(\ref{ope2})
can be read directly from the $N=2$ minimal CFT results in \cite{Marian}
using the definition (\ref{l-m}).
The only difference is the normalization constant of the conformal blocks
due to the Clebsch-Gordan coefficients in the OPEs of the (NS) fields
which was not accounted there.
With this normalization the conformal blocks are given by:
\beq
{1\over{1-\val b}}F\left({\val+\bar\val\over b}-{2\over b^2},
1-{1\over b^2};{\val+\bar\val\over b}-{1\over b^2}+1;\eta\right)
\eeq
for the first term in the RHS of Eq.(\ref{ope2}) and
\beq
{1\over\bar\val b}F\left(-{\val+\bar\val\over b},1-{1\over b^2};1+{1\over b^2}
-{\val+\bar\val\over b};\eta\right)
\eeq
for the second one.
The structure constant $C_-^{(NS)}$ can also be extracted from \cite{Marian}:
\beq
C_-^{(NS)}=-{\tilde\mu}'
{({\bar\val}b)^2\Ga(-{\val+\bar\val\over b})\Ga({\val+\bar\val\over b}-{1\over b^2})\over
\Ga(1+{1\over b^2}-{\val+\bar\val\over b})\Ga(1+{\val+\bar\val\over b})},
\eeq
where ${\tilde\mu}'$ is given by ${\tilde\mu}$, the cosmological constant of
the dual $N=2$ theory,
\beq
{\tilde\mu}'=4\pi{\tilde\mu}\gamma(1+b^{-2})b^{-4}.
\eeq
With all these ingredients we are able to write down new functional
equations for the one-point functions of non-neutral primary fields:
\beaq
{\cal R}(-1/b,0)U^{NS}(\val,\bar\val)&=&
{\Ga(-{2\over b^2})\Ga({\val+\bar\val\over b}-{1\over b^2}+1)\over{
(1-\val b)\Ga(1-{1\over b^2})\Ga({\val+\bar\val\over b}-{2\over b^2})}}
U^{NS}\left(\val-{1\over b},\bar\val\right)\nonumber\\
&-&{\tilde\mu}'
{(\bar\val b)\Ga(-{2\over b^2})\Ga({\val+\bar\val\over b}-{1\over b^2})\over{
\Ga(1-{1\over{b^2}})\Ga(1+{\val+\bar\val\over b})}}
U^{NS}\left(\val,\bar\val+{1\over b}\right),
\label{funeqNiii}
\eeaq
Again, we do not know the bulk-boundary structure constant ${\cal R}(-1/b,0)$.

The (R) sector can be treated in exactly the same way.
Similarly to (\ref{Nlm}), we can identify
\beq
R^l_{m,1}=\sigma^{+}e^{-{l+m+1\over 2b}\phi^{+}-{l-m-1\over 2b}\phi^{-}}.
\eeq
The OPE is given by
\beq
N_{-1/b,0}R^{(+)}_{\val,\bar\val}=R^{(+)}_{\val-1/b,\bar\val}
+ C_{-}^{(R)}R^{(+)}_{\val,\bar\val +1/b}.
\label{ope3}
\eeq

The conformal blocks can be computed as before
\beq
{1\over{3/2-\val b}}F\left({\val+\bar\val\over b}-{2\over b^2},
1-{1\over b^2};{\val+\bar\val\over b}-{1\over b^2}+1;\eta\right)
\eeq
for the first term in the RHS of Eq.(\ref{ope3}) and
\beq
{1\over{\bar\val b+1/2}}F\left(-{\val+\bar\val\over b},
1-{1\over b^2};1+{1\over b^2}-{\val+\bar\val\over b};\eta\right)
\eeq
for the second one.
The structure constant $C_-^{(R)}$ is given by
\beq
C_-^{(R)}=-{\tilde\mu}'
{(\bar\val b+{1\over{2}})^2\Ga(-{\val+\bar\val\over b})\Ga({\val+\bar\val\over b}
-{1\over b^2})\over
\Ga(1+{1\over b^2}-{\val+\bar\val\over b})\Ga(1+{\val+\bar\val\over b})}.
\eeq
With these, we find the functional equation for the (R) field:
\beaq
{\cal R}(-1/b,0)U^{R}(\val,\bar\val)&=&
{\Ga(-{2\over b^2})\Ga({\val+\bar\val\over b}-{1\over b^2}+1)\over{
({3\over{2}}-\val b)\Ga(1-{1\over b^2})\Ga({\val+\bar\val\over b}-{2\over b^2})}}
U^{R}\left(\val-{1\over b},\bar\val\right)\nonumber\\
&-&{\tilde\mu}'
{(\bar\val b+{1\over{2}})\Ga(-{2\over b^2})\Ga({\val+\bar\val\over b}-{1\over b^2})
\over{\Ga(1-{1\over{b^2}})\Ga(1+{\val+\bar\val\over b})}}
U^{R}\left(\val,\bar\val+{1\over b}\right).
\label{funeqRiii}
\eeaq

One can check that these equations are consistent with the functional equations
(\ref{funceqdual}) derived in the sect.3.3 for neutral fields with
$\al={\bar\al}$.
This justifies the method of analytic continuation used in this subsection.
From now on, we will use Eqs.(\ref{funeqNiii}) and (\ref{funeqRiii}) instead of
Eq.(\ref{funceqdual}) since they are applicable even to non-neutral fields.

\subsection{Solutions}

In our previous paper \cite{ASY}, we have derived the one-point functions
on the FZZT BC based on the modular bootstrap method and confirmed their validity
using a single functional equation.
Now it is possible to solve the set of functional equations to determine
the one-point functions completely.
They are given by
\beaq
U^{NS}_{s}(\al,{\bar\al})
&=&\left(\pi\mu\right)^{-{\al+{\bar\al}\over{b}}+{1\over{b^2}}}
{\Gamma\left(1+{\al+{\bar\al}\over{b}}-{1\over{b^2}}\right)
\Gamma\left(b(\al+{\bar\al})-1\right)
\over{\Gamma\left(\al b\right)\Gamma\left({\bar\al}b\right)}}\nonumber\\
&\times&\cosh\left[2\pi s\left(\al+{\bar\al}-{1\over{b}}\right)\right],
\label{unss}\\
U^{R}_{s}(\al,{\bar\al})
&=&\left(\pi\mu\right)^{-{\al+{\bar\al}\over{b}}+{1\over{b^2}}}
{\Gamma\left(1+{\al+{\bar\al}\over{b}}-{1\over{b^2}}\right)
\Gamma\left(b(\al+{\bar\al})-1\right)
\over{\Gamma\left(\al b-{1\over{2}}\right)
\Gamma\left({\bar\al}b+{1\over{2}}\right)}}\nonumber\\
&\times&\cosh\left[2\pi s\left(\al+{\bar\al}-{1\over{b}}\right)\right].
\label{urs}
\eeaq
The continuous parameter $s$ is related nonperturbatively to the
boundary cosmological constant $\mu_{B}$ in Eq.(\ref{baction}) by
\beq
{\overline\mu_B}^2 ={\mu b^2\over{2}}\cosh(2\pi sb)
\label{boundparam}
\eeq
along with the relation between the two cosmological constants
\beq
4\pi{\tilde\mu}\gamma(1+b^{-2})=(\pi\mu)^{2/b^2}.
\eeq
These results match perfectly with those of the modular bootstrap \cite{ASY}.

\section{ZZ-Branes}

In this section we are interested in the $N=2$ SLFT on Lobachevskiy plane
or pseudosphere which is the geometry of the infinite constant negative
curvature surface.
Using previous conformal bootstraps, we derive and solve
nonlinear functional equations which can provide discrete BCs.

\subsection{Pseudosphere geometry}

The classical equations of motion for the $N=2$ SLFT can be derived from
the Lagrangian (\ref{N2L}):
\beaq
\partial\bar\partial\phi^{\pm}&=& \pi\mu b^3\left[\pi\mu
e^{b(\phi^{+}+\phi^{-})}+i\psi^{\pm}{\bar\psi}^{\pm}e^{b\phi^{\mp}}\right]\\
\partial{\bar\psi}^{\pm}&=& i\pi \mu b^2 e^{b\phi^{\pm}}\psi^{\mp},\qquad
\bar\partial\psi^{\pm}= -i\pi \mu b^2 e^{b\phi^{\pm}}{\bar\psi}^{\mp}.
\eeaq
Assuming that the fermionic fields vanish in the classical limit, we can
solve the bosonic fields classically
\beq
e^{\varphi(z)}={4R^2\over{(1-|z|^2)^2}},
\eeq
where $\varphi=b(\phi^{+}+\phi^{-})$ and $R^{-2}=4\pi^2\mu^2 b^4$.
The parameter $R$ is interpreted as the radius of the pseudosphere
in which the points at the circle $|z|=1$ are infinitely
far away from any internal point.
This circle can be interpreted as the ``boundary'' of the pseudosphere.

This boundary has a different class of BCs, which are classified by integers
whose interpretation is not clear yet.
For the $N=2$ SLFT, we will call the discrete BCs as ZZ-branes following \cite{ZZ}
and show that these correspond to the degenerate fields of the $N=2$ SLFT.

\subsection{Conformal bootstrap equations on pseudosphere}

In sect.3, we have started with two-point correlation functions on a half plane.
As the two fields approach on the boundary, the degenerate field is expanded into
the boundary operators.
On pseudosphere geometry, as they approach on the boundary $\eta\rightarrow 1$,
the distance between the two points become infinite due to the singular metric.
This means that the two-point function is factorized into a product of two
one-point functions.
For example, the two-point function in Eq.(\ref{twopt}) becomes
\beq
G_{\al}(\xi,\xi')={|\xi'-\bar\xi'|^{2\Delta^{NS}_\val-2\Delta^{NS}_{-b/2}}\over
|\xi-\bar\xi'|^{4\Delta^{NS}_\val}}U^{NS}(-b/2)U^{NS}(\val)
{\tilde{\cal F}}^{NS}_3 (\eta).
\eeq
Meanwhile, the computation of the other channel where we take the OPE of the two
fields first is identically same as in sect.3.
Comparing these two results, we can obtain the following
nonlinear functional equations for $U(\val)$:
\beaq
{\cal C}_1U^{NS}(-b/2)U^{NS}(\val)&=&{\Ga(\val b-{b^2\over 2})
\Ga(\val b+{1\over 2})\over \Ga(\val b)\Ga(\val b-{b^2\over 2}-{1\over 2})}
U^{NS}\left(\val-{b\over 2}\right)\nonumber\\
&+&2^{-2-2b^2}\pi^2 b^4\mu^2{\Ga(\val b-{1\over 2})\Ga(\val b+{b^2\over 2})
\over \Ga(\val b)\Ga(\val b+{b^2\over 2}+{1\over 2})}
U^{NS}\left(\val +{b\over 2}\right)
\label{psi}
\eeaq
with
\beq
{\cal C}_1={\sqrt{\pi}\Ga(-{b^2\over 2})\over{
\Ga(-1)\Ga(-{b^2\over 2}-{1\over 2})}}.
\label{c1}
\eeq

Similarly we can derive functional equations corresponding to
Eqs.(\ref{funeqNiii}) and (\ref{funeqRiii}):
\beaq
{\cal C}_2{\tilde U}^{NS}(-1/b,0)U^{NS}(\val,\bar\val)&=&
{\Ga({\val+\bar\val\over b}-{1\over b^2}+1)\over{
(1-\val b)\Ga({\val+\bar\val\over b}-{2\over b^2})}}
U^{NS}\left(\val-{1\over b},\bar\val\right)\nonumber\\
&-&{\tilde\mu}'
{(\bar\alpha b)\Ga({\val+\bar\val\over b}-{1\over b^2})\over{
\Ga(1+{\val+\bar\val\over b})}}
U^{NS}\left(\val,\bar\val+{1\over b}\right),
\label{psiv}\\
{\cal C}_2{\tilde U}^{NS}(-1/b,0)U^{R}(\val,\bar\val)&=&
{\Ga({\val+\bar\val\over b}-{1\over b^2}+1)\over{
({3\over{2}}-\val b)\Ga({\val+\bar\val\over b}-{2\over b^2})}}
U^{R}\left(\val-{1\over b},\bar\val\right)\nonumber\\
&-&{\tilde\mu}'
{(\bar\val b+{1\over{2}})\Ga({\val+\bar\val\over b}-{1\over b^2})
\over{\Ga(1+{\val+\bar\val\over b})}}
U^{R}\left(\val,\bar\val+{1\over b}\right)
\label{psv}
\eeaq
with ${\cal C}_2=\Ga(1-{1\over{b^2}})/\Ga(-{2\over{b^2}})$.
Here we have denoted one-point functions of the class-II degenerate field in
terms of ${\tilde U}^{NS}$ since they are in principle different from the
one-point functions of general fields.

Notice that ${\cal C}_1$ contains $\Gamma(-1)$ in Eq.(\ref{c1}) which arises
from singular monodromy transformation.
We can remove this singular factor by redefining $U^{NS(R)}/\Gamma(-1)\to U^{NS(R)}$.
Notice that this redefinition does not change Eqs.(\ref{psiv}) and (\ref{psv})
since they are linear in $U^{NS(R)}$ if assuming that ${\tilde U}^{NS}$ is regular.

\subsection{Solutions}

The solutions to these equations can be expressed in terms of two integers
$m,n\ge 1$ as follows:
\beaq
U^{NS}_{mn}(\val,\bar\al)&=&{\cal N}_{mn}(\pi\mu)^{-{\al+{\bar\al}\over{b}}}
{\Ga({\al+{\bar\al}\over{b}}-{1\over b^2}+1)\Ga(b(\al+{\bar\al})-1)\over
\Ga(\al b)\Ga({\bar\al} b)}\nonumber\\
&\times&\sin\left[{\pi m\over b}\left(\al+{\bar\al}-{1\over b}\right)\right]
\sin\left[\pi nb\left(\al+{\bar\al}-{1\over b}\right)\right]
\label{ZZNS}
\\
U^{R}_{mn}(\val,\bar\al)&=&{\cal N}_{mn}(\pi\mu)^{-{\al+{\bar\al}\over{b}}}
{\Ga({\al+{\bar\al}\over{b}}-{1\over b^2}+1)\Ga(b(\al+{\bar\al})-1)\over
\Ga(\al b-1/2)\Ga({\bar\al} b+1/2)}\nonumber\\
&\times&\sin\left[{\pi m\over b}\left(\al+{\bar\al}-{1\over b}\right)\right]
\sin\left[\pi nb\left(\al+{\bar\al}-{1\over b}\right)\right],
\label{ZZR}
\eeaq
with the normalization factors given by
\beq
{\cal N}_{mn}=(-1)^n {4b^2\over{\Ga(-1/b^2)}}
{\cot(\pi nb^2)\over{\sin(\pi m/b^2)}}.
\eeq
This class of solutions will be associated with conformal BCs corresponding
to the class-I neutral degenerate fields.
It turns out that the conformal bootstrap equations do not allow
discrete BCs corresponding to non-neutral degenerate fields.
One possible explanation is that non-neutral BCs will introduce a boundary
field which will not produce the identity operator when fused with bulk
degenerate fields as they approach the boundary.

It is interesting to notice that the following one-point functions
\beaq
U^{NS}_m(\alpha,\bar{\alpha})&=&{\cal N}_m(\pi\mu)^{-\frac{\alpha+
\bar{\alpha}}{b}}\frac{\Gamma(1-\alpha b)\Gamma(1-\bar{\alpha}b)}
{\Gamma(-\frac{\alpha+\bar{\alpha}}{b}+\frac{1}{b^2})\Gamma(2-b(\alpha
+\bar{\alpha}))}\nonumber\\
&\times&\frac{\sin\left[\frac{\pi m}{b}(\alpha+\bar{\alpha}-\frac{1}{b})\right]}
{\sin\left[\frac{\pi}{b}(\alpha+\bar{\alpha}-\frac{1}{b})\right]}
\label{UNSm}\\
U^R_m(\alpha,\bar{\alpha})&=&{\cal N}_m(\pi\mu)^{-\frac{\alpha+\bar{\alpha}}{b}}
\frac{\Gamma(\frac{3}{2}-\alpha b)\Gamma(\frac{1}{2}-\bar{\alpha}b)}
{\Gamma(-\frac{\alpha+\bar{\alpha}}{b}+\frac{1}{b^2})\Gamma(2-b(\alpha
+\bar{\alpha}))}\nonumber\\
&\times&\frac{\sin\left[\frac{\pi m}{b}(\alpha+\bar{\alpha}-\frac{1}{b})\right]}
{\sin\left[\frac{\pi}{b}(\alpha+\bar{\alpha}-\frac{1}{b})\right]}
\label{URm}\\
{\cal N}_m&=&\frac{\pi}{\Gamma(-\frac{1}{b^2}+1)}\frac{1}
{\sin(\frac{\pi m}{b^2})}
\eeaq
satisfy Eqs.(\ref{psiv}) and (\ref{psv}).
Although these do not satisfy Eq.(\ref{psi}), hence not complete solutions,
this class of solutions turns out to be consistent with modular bootstrap equations and
we will show that they correspond to the class-III BCs.

\section{Modular Bootstrap}

In this section we derive the modular bootstrap equations based on
the modular properties of degenerate characters.
We derive the boundary amplitudes which are consistent with the one-point
functions derived before.

\subsection{Characters of general primary fields}

The character of a CFT is defined by the trace over
all the conformal states built on a specific primary state
\beq
\chi_h(q,y,t)=e^{2\pi ik t}{\rm Tr}\left[q^{L_0-c/24}y^{J_0}\right],
\eeq
with $k=c/3$.
Since the primary fields with general $\al,{\bar\al}$ of
the $N=2$ SLFT have no null states,
the characters can be obtained by simply summing up all the descendant states.
For these primary fields, it is more convenient to use the real parameters
$P,\omega$ to denote them using Eqs.(\ref{u1charge}) and (\ref{momentum}).
The (NS) character can be computed as
\beq
\chi^{NS}_{[P,\omega]}(q,y,t)
=e^{2\pi i kt} q^{P^2+b^2\omega^2/4} y^{\omega}
{\theta_{00}(q,y)\over{\eta(q)^3}},
\label{nschi}
\eeq
where we have introduced standard elliptic functions
\[
\eta(q)=q^{1/24}\prod_{n=1}^{\infty}(1-q^n),\quad
\theta_{00}(q,y)=\prod_{n=1}^{\infty}\left[
(1-q^n)(1+yq^{n-1/2})(1+y^{-1}q^{n-1/2})\right].
\]

For the conformal BCs of super-CFTs, one needs to consider
characters and associated Ishibashi states of the
($\widetilde{\rm NS}$) sectors \cite{Nepomechie}.
The ($\widetilde{\rm NS}$) characters are defined by
\beq
\chi^{\widetilde{\rm NS}}_h(q,y,t)=
e^{2\pi i kt}{\rm Tr}\left[(-1)^Fq^{L_0-c/24}y^{J_0}\right].
\eeq
For a primary field $N_{\al{\overline\al}}$, $(-1)^F$ term
contributes $-1$ for those descendants with odd number of $G^{\pm}_{-r}$.
This can be efficiently incorporated into the character formula
by shifting $y\to -y$ in the product.
Therefore, the ($\widetilde{\rm NS}$) character is given by
\beq
\chi^{\widetilde{NS}}_{[P,\omega]}(q,y,t)=e^{2\pi i kt}
q^{P^2+b^2\omega^2/4} y^{\omega} {\theta_{00}(q,-y)\over{\eta(q)^3}}.
\label{nstildechi}
\eeq

The character of a (R) primary field $R_{[P,\omega]}^{(\ep)}$ is given by
\beq
\chi^{R}_{[P,\omega,\epsilon]}(q,y,t)=
e^{2\pi i kt} q^{P^2+b^2\omega^2/4} y^{\omega}
{\theta_{10}(q,y)\over{\eta(q)^3}},\label{rchi}
\eeq
where we introduce another elliptic function
\beq
\theta_{10}(q,y)=(y^{1/2}+y^{-1/2})q^{1/8}\prod_{n=1}^{\infty}\left[
(1-q^n)(1+yq^{n})(1+y^{-1}q^{n})\right].
\eeq

\subsection{Characters of degenerate fields}

In this subsection, we will consider only the (NS) fields which will be
used later.
The (R) characters can be similarly computed.
Let us start with a class-I (NS) degenerate field $N_{m,n}^{\omega}$.
As claimed in Eq.(\ref{nullone}), this field has a null state.
Therefore, the character is given by
\beq
\chi^{NS}_{mn\omega}(q,y,t)=e^{2\pi ikt}
\left[q^{-\frac{1}{4}(\frac{m}{b}+nb)^2}-q^{-\frac{1}{4}(\frac{m}{b}-nb)^2}\right]
q^{b^2\omega^2/4}y^{\omega}\frac{\theta_{00}(q,y)}{\eta(q)^3}.
\label{NSde1ch}
\eeq

The characters of the class-II (NS) degenerate fields,
$N_{m}^{\omega}$ and ${\tilde N}_{m}^{\omega}$, are rather complicated
due to the infinite null states structure.
One should add and subtract contributions of these null states
infinitely.
The character of a class-IIA degenerate field is given by
\beaq
\chi^{NS}_{m\omega}(q,y,t)&=&e^{2\pi ikt}\frac{\theta_{00}(q,y)}{\eta(q)^3}
\sum_{j=0}^{\infty}(-1)^j q^{-m^2/4b^2}(yq^{m/2})^{\omega+j}\\
&=&e^{2\pi ikt}
{y^{\omega}q^{-m^2/4b^2+m\omega/2}\over{1+yq^{m/2}}}
\frac{\theta_{00}(q,y)}{\eta(q)^3}
\eeaq
and similarly for a class-IIB:
\beq
\chi^{NS}_{m\omega}(q,y,t)=e^{2\pi ikt}
{y^{\omega}q^{-m^2/4b^2-m\omega/2}\over{1+y^{-1}q^{m/2}}}
\frac{\theta_{00}(q,y)}{\eta(q)^3}.
\label{NSde2ach}
\eeq

One should be more careful for the neutral class-III degenerate fields.
There are two infinite semi-chains of null states as expressed in
Eq.(\ref{neutralseq}).
Adding all these states, one can find the character as follows:
\beaq
\chi^{NS}_{m}(q,y,t)&=&e^{2\pi ikt}\frac{\theta_{00}(q,y)}{\eta(q)^3}
\left[\sum_{j=0}^{\infty}q^{-m^2/4b^2}(-yq^{m/2})^j+
\sum_{j=1}^{\infty}q^{-m^2/4b^2}(-y^{-1}q^{m/2})^j\right]\nonumber\\
&=&e^{2\pi ikt}\frac{q^{-m^2/4b^2}(1-q^m)}{(1+yq^{m/2})(1+y^{-1}q^{m/2})}
\frac{\theta_{00}(q,y)}{\eta(q)^3}.
\label{NSde2cch}
\eeaq
When $m=1$, this character is the same as that of the identity operator
as expected.

\subsection{Modular transformations}

The modular transformation of the class-I character can be easily found as
\beq
\chi^{NS}_{mn\omega}(q',y',t')=
2b\int_{-\infty}^{\infty}dP\int_{-\infty}^{\infty}d\omega'
\sinh(2\pi mP/b)\sinh(2\pi nbP)e^{-\pi ib^2\omega\omega'}
\chi^{NS}_{[P,\omega']}(q,y,t).
\label{NSde1modtrans}
\eeq
Here we have used $q',y',t'$ for the $S$-modular transformed parameters and
$\omega'$ as the $U(1)$ charge of a general primary field to distinguish it
from that of the degenerate field $\omega$.

The modular transformation of the class-II characters
can be derived by the method of \cite{Miki}:
\beaq
\chi^{NS}_{m\om}(q',y',t')&=&\frac{b}{2}\int_{-\infty}^{\infty}dP
\int_{-\infty}^{\infty}d\omega'
\Big[\frac{e^{-\pi ib^2\om\omega'}\cosh[2\pi bP(\om\mp 1\mp\frac{m}{b^2})]}
{2\cosh(\pi bP+\frac{\pi ib^2\omega'}{2})\cosh(\pi bP-\frac{\pi ib^2\omega'}{2})}
\nonumber\\
&+&\frac{e^{-\pi ib^2(\om\mp 1)\omega'}\cosh[2\pi bP(\om\mp\frac{m}{b^2})]}
{2\cosh(\pi bP+\frac{\pi ib^2\omega'}{2})\cosh(\pi bP-\frac{\pi ib^2\omega'}{2})}
\Big]\chi^{NS}_{[P,\omega']}(q,y,t)
\nonumber\\
&\pm&i\sum_{r\in {\mathbf Z}+{1\over{2}}}\int_{0}^{1}d\la
e^{-i\pi (\pm m\la+2\omega r)} {\tilde \chi}^{NS}_{r\la}(q,y,t),
\label{NSde2amodtrans}
\eeaq
where the upper (lower) sign denotes the class-IIA (IIB), respectively.

For the class-III degenerate fields, the transformation is given by
\beaq
\chi^{NS}_m(q',y',t')&=&\frac{b}{2}\int_{-\infty}^{\infty}dP
\int_{-\infty}^{\infty}d\omega'
\frac{\sinh(2\pi mP/b)\sinh(2\pi bP)}
{\cosh\left(\pi bP+\frac{\pi ib^2\omega'}{2}\right)
\cosh\left(\pi bP-\frac{\pi ib^2\omega'}{2}\right)}
\chi^{NS}_{[P,\omega']}(q,y,t)
\nonumber\\
&+&2\sum_{r\in{\mathbf Z}+{1\over{2}}}\int_{0}^{1}d\la
\sin(\pi m\la) {\tilde \chi}^{NS}_{r\la}(q,y,t).
\label{NSde2cmodtrans}
\eeaq
Here we have defined a spectral flow of the class-IIA character
\beq
{\tilde \chi}^{NS}_{r\la}(q,y,t)\equiv e^{2\pi ikt}
\frac{y^{2r/b^2+\la}q^{r^2/b^2+r\la}}{1+yq^{r}}
\frac{\theta_{00}(q,y)}{\eta(q)^3}.
\eeq

\subsection{Conformal boundary conditions}

\subsubsection{Vacuum BC}

According to Cardy's formalism, one can associate
a conformal BC with each primary state \cite{Cardy}.
Among the conformal BCs of the $N=2$ SLFT, we concentrate on those associated
with the degenerate fields.
Following the modular bootstrap formulation, we can compute a boundary amplitude
which is the inner product between
the Ishibashi state of a primary state and the conformal boundary state.
As usual, we start with the `vacuum' BC amplitude \cite{EguSug,ASY}:
\beq
\Psi_{\mathbf 0}^{NS}(P,\omega){\Psi_{\mathbf 0}^{NS}}^{\dagger}(P,\omega)
={\mathbf S}_{NS}(P,\omega),
\label{nsvac}
\eeq
where the boundary amplitude is defined by
\beq
\Psi_{\mathbf 0}^{NS}(P,\omega)=\langle {\mathbf 0}\vert N_{[P,\omega]}
\rangle\rangle
\eeq
and the modular $S$-matrix element ${\mathbf S}_{NS}(P,\omega)$ is given by
Eq.(\ref{NSde2cmodtrans}) with $m=1$.

Since ${\Psi_{\mathbf 0}^{NS}}^{\dagger}(P,\omega)
=\Psi_{\mathbf 0}^{NS}(-P,\omega)$, one can solve this up to some
unknown constant as follows:
\beq
\Psi^{NS}_{\mathbf 0}(P,\omega)=\sqrt{{b^3\over{2}}}
\left(\pi\mu\right)^{-{2iP\over{b}}}
{\Gamma\left({1\over{2}}-ibP+{b^2\omega\over{2}}\right)
\Gamma\left({1\over{2}}-ibP-{b^2\omega\over{2}}\right)\over{
\Gamma\left(-{2iP\over{b}}\right)\Gamma\left(1-2ibP\right)}}.
\label{NSvac}
\eeq

Similarly, the vacuum boundary amplitude for the (R) Ishibashi state is
given by \cite{ASY}:
\beq
\Psi^{R}_{\mathbf 0}(P,\omega)=-i\sqrt{{b^3\over{2}}}
\left(\pi\mu\right)^{-{2iP\over{b}}}
{\Gamma\left(-ibP+{b^2\omega\over{2}}\right)
\Gamma\left(1-ibP-{b^2\omega\over{2}}\right)\over{
\Gamma\left(-{2iP\over{b}}\right)\Gamma\left(1-2ibP\right)}}.
\eeq

\subsubsection{Class-I BCs}

Now we impose the vacuum BC on one boundary and discrete BCs associated with
the class-I degenerate fields on the other.
Following the Cardy formalism, we can get the relation
\beq
\chi^{NS}_{mn\omega}(q',y',t')
=\int_{-\infty}^{\infty}dP\int_{-\infty}^{\infty}d\omega'
\Psi^{NS}_{mn\omega}(P,\omega')\Psi^{NS\dag}_{\bf 0}(P,\omega')
\chi^{NS}_{[P,\omega']}(q,y,t).
\label{NSde1PsiPsi}
\eeq
Here, the boundary amplitudes are defined by an inner product between the
the boundary state $\vert m,n,\omega\rangle$ and the Ishibashi state
\beq
\Psi_{mn\omega}^{NS}(P,\omega)=\langle m,n,\omega\vert N_{[P,\omega]}\rangle\rangle.
\eeq
Comparing this with Eq.(\ref{NSde1modtrans}), we obtain
\beq
\Psi^{NS}_{mn\omega}(P,\omega')\Psi^{NS\dag}_{\bf 0}(P,\omega')=
2b\sinh(2\pi mP/b)\sinh(2\pi nbP)e^{-\pi ib^2\omega\omega'}.
\label{NSde1Psi}
\eeq
We can find the boundary amplitude from Eq.(\ref{NSvac})
\beaq
\Psi^{NS}_{mn\omega}(P,\omega')&=&
\sqrt{{8\over{b}}}
\left(\pi\mu\right)^{-{2iP\over{b}}}
{\Gamma\left({2iP\over{b}}\right)\Gamma\left(1+2ibP\right)
\over{\Gamma\left({1\over{2}}+ibP+{b^2\omega\over{2}}\right)
\Gamma\left({1\over{2}}+ibP-{b^2\omega\over{2}}\right)}}\nonumber\\
&\times&\sinh(2\pi mP/b)\sinh(2\pi nbP)
e^{-\pi ib^2\omega\omega'}.
\label{NSIres}
\eeaq

It is remarkable that this solution coincides with Eq.(\ref{ZZNS}),
the ZZ-brane solution with BC $(m,n,\omega=0)$.
This provides a most important consistency check between the conformal
and modular bootstraps.

Now we impose the class-I discrete BCs on both boundaries.
The partition function is expressed by
\beq
Z^{NS}_{(mn\om)(m'n'\om')}(q',y',t')=
\int_{-\infty}^{\infty}dP\int_{-\infty}^{\infty}d\omega''
\chi^{NS}_{[P,\omega'']}(q,y,t)\Psi^{NS}_{mn\om}(P,\omega'')
\Psi^{NS\dag}_{m'n'\om'}(P,\omega'').
\eeq
Inserting Eq.(\ref{NSIres}) into this, we find
\beq
Z^{NS}_{(mn\om)(m'n'\om')}=\sum_{k=|m-m'|+1}^{m+m'-1}\sum_{l=|n-n'|+1}^{n+n'-1}
[\chi^{NS}_{k,l-1,\om+\om'}+\chi^{NS}_{k,l+1,\om+\om'}
+\chi^{NS}_{k,l,\om+\om'+1}+\chi^{NS}_{k,l,\om+\om'-1}],
\label{NSde1Z}
\eeq
where we have omitted the modular parameters for simplicity.
From Eq.(\ref{NSde1Z}) we can read off the fusion rules of the class-I
degenerate fields.
In particular, the OPE between neutral fields are given by
\beq
N_{mn}^{0}N_{m'n'}^{0}=\sum_{k=|m-m'|+1}^{m+m'-1}
\sum_{l=|n-n'|+1}^{n+n'-1}\left[N_{k,l-1}^{0}+N_{k,l+1}^{0}+
N_{k,l}^{1}+N_{k,l-1}^{-1}\right].
\eeq
Notice that this fusion rule can not be applicable to $n=n'$ case
where $l=1,\ldots,2n-1$.
If $l=1$, the field $N_{k,1}^{1}$ is not in class-I as mentioned before and
the relation (\ref{NSde1Z}) is not valid.
This explains why the OPE of two identical class-I degenerate fields includes
identity field which is not in class-I but in class-III.

So far we have considered the discrete BCs on both boundaries.
It is interesting to consider a mixed BC, namely a class-I BC on one
boundary and the continuous (FZZT) BC \cite{ASY} on the other.
In this case, the partition function can be written as
\beq
Z^{NS}_{(mn\om)s}(q',y',t')=
\int_{-\infty}^{\infty}dP\int_{-\infty}^{\infty}d\omega'
\chi^{NS}_{[P,\omega']}(q,y,t)\Psi^{NS}_{mn\om}(P,\omega')
\Psi^{NS\dag}_{s}(P,\omega'),
\eeq
with
\beq
\Psi^{NS}_{\mathbf s}(P,\omega')=\sqrt{2b^3}\left(\pi\mu\right)^{-{2iP\over{b}}}
{\Gamma\left(1+{2iP\over{b}}\right)\Gamma\left(2ibP\right)
\cos(4\pi s P)\over{\Gamma\left({1\over{2}}+ibP+{b^2\omega'\over{2}}\right)
\Gamma\left({1\over{2}}+ibP-{b^2\omega'\over{2}}\right)}},
\label{ampns}
\eeq
which is in fact Eq.(\ref{unss}) up to a proportional constant.
Inserting Eq.(\ref{NSIres}) into this, we obtain the following result:
\beq
Z^{NS}_{(mn\om)s}=\sum_{k=1-m,2}^{m-1}\sum_{l=1-n,2}^{n-1}
[\chi^{NS}_{[P_{-k,-l-1},\om]}+\chi^{NS}_{[P_{k,l+1},\om]}+
\chi^{NS}_{[P_{k,l},\om+1]}+\chi^{NS}_{[P_{k,l},\om-1]}],
\label{NSmixed}
\eeq
where we have omitted the modular parameters for simplicity and introduced
a momentum variable
\beq
P_{k,l}=s+{i\over{2}}\left({k\over{b}}+lb\right).
\eeq

It is more illustrative to consider the most simple case, namely, $m=n=1,\omega=0$.
This BC is associated with the class-I degenerate field $N_{-b/2}$
which we have considered in sect.3.2.
The above equation is simplified to
\beq
Z^{NS}_{(110)s}=\chi^{NS}_{[s-ib/2,0]}+\chi^{NS}_{[s+ib/2,0]}+
\chi^{NS}_{[s,1]}+\chi^{NS}_{[s,-1]}.
\label{NSmixedi}
\eeq
In terms of $\alpha,{\bar\alpha}$, one can easily check that these are
characters of the operators appearing in the OPE (\ref{ope}).
This provides another consistency check for our results.

\subsubsection{Class-II BCs}

For the class-II and class-III (neutral) BCs,
there are two types of Ishibashi states flowing in the bulk.
One is associated with the continuous state $\vert N_{[P,\omega']}\rangle\rangle$
and the other with class-II degenerate fields and their spectral flows.
We denote this Ishibashi state by $\vert r,\la\rangle\rangle$.
The appearance of this state can be understood from the modular transformations,
Eq.(\ref{NSde2amodtrans}).

If we denote the class-II boundary state $\vert m,\omega\rangle$,
we can define the following boundary amplitudes as inner products
between the boundary state and the Ishibashi states
\beq
\Psi_{m\omega}^{NS}(P,\omega')=
\langle m,\omega\vert N_{[P,\omega']}\rangle\rangle,\qquad
\Phi_{m\omega}^{NS}(r,\la)=\langle m,\omega\vert r,\la\rangle\rangle.
\eeq
Using these, one can express the partition function with
the vacuum BC on one side and a class-II BC on the other boundary
\beaq
\chi^{NS}_{m\omega}(q',y',t')
&=&\int_{-\infty}^{\infty}dP\int_{-\infty}^{\infty}d\omega'
\Psi^{NS}_{m\omega}(P,\omega')\Psi^{NS\dag}_{\bf 0}(P,\omega')
\chi^{NS}_{[P,\omega']}(q,y,t)\nonumber\\
&+&\sum_{r\in Z+{1\over{2}}}\int_{0}^{1}d\la
\Phi_{m\omega}^{NS}(r,\la)\Phi_{1}^{NS\dagger}(r,\la)
{\tilde \chi}^{NS}_{r\la}(q,y,t).
\label{NSde2PsiPsi}
\eeaq
Comparing this with Eq.(\ref{NSde2amodtrans}), we obtain
\beq
\Psi^{NS}_{m\omega}(P,\omega')\Psi^{NS\dag}_{\bf 0}(P,\omega')=
S_{m\omega}(P,\omega'),
\label{NSde2Psi}
\eeq
where $S_{m\omega}(P,\omega')$ is the modular $S$-matrix component in
Eq.(\ref{NSde2amodtrans}).
From this, one can solve for $\Psi^{NS}_{m\omega}(P,\omega')$.
Instead of presenting details for this case, we will analyze
more interesting case, namely the neutral ($\omega=0$) class-III BCs.

\subsubsection{Class-III BCs}

For a class-III (neutral) boundary state $\vert m\rangle$,
we can define two boundary amplitudes
\beq
\Psi_{m}^{NS}(P,\omega)=\langle m\vert N_{[P,\omega]}\rangle\rangle,\qquad
\Phi_{m}^{NS}(r,\la)=\langle m\vert r,\la\rangle\rangle
\eeq
due to the two Ishibashi states.
Imposing this BC on one side and the vacuum BC on the other, we can find
\beaq
\chi^{NS}_m(q',y',t')&=&\int_{-\infty}^{\infty}dP
\int_{-\infty}^{\infty}d\omega'
\Psi^{NS}_{m}(P,\omega')\Psi^{NS\dag}_{\bf 0}(P,\omega')
\nonumber\\
&+&\sum_{r\in Z+{1\over{2}}}\int_{0}^{1}d\la
\Phi_{m}^{NS}(r,\la)\Phi_{1}^{NS\dagger}(r,\la){\tilde \chi}^{NS}_{r\la}(q,y,t).
\label{NSde2cPsiPsi}
\eeaq
Comparing with Eq.(\ref{NSde2cmodtrans}), we obtain
\beq
\Psi^{NS}_m(P,\omega)={\Psi_{\mathbf 0}^{NS}}(P,\omega)
\frac{\sinh(2\pi mP/b)}{\sinh(2\pi P/b)},
\label{NSde2cPsi}
\eeq
and
\beq
\Phi^{NS}_m(r,\la)={2\sin(m\pi\la)\over{\sqrt{2\sin(\pi\la)}}}.
\label{disampIII}
\eeq
The solution (\ref{NSde2cPsi}) coincides with the one-point function (\ref{UNSm}).
Imposing these BCs on both boundaries, the partition function is given by
\beaq
Z^{NS}_{mm'}(q',y',t')&=&\int_{-\infty}^{\infty}dP\int_{-\infty}^{\infty}d\omega
\chi^{NS}_{[P,\omega]}(q,y,t)\Psi^{NS}_{m}(P,\omega)\Psi^{NS\dag}_{m'}(P,\omega)
\nonumber\\
&+&\sum_{r\in Z+{1\over{2}}}\int_{0}^{1}d\la
\Phi_{m}^{NS}(r,\la)\Phi_{m'}^{NS\dagger}(r,\la){\tilde \chi}^{NS}_{r\la}(q,y,t).
\label{NSde2cZ}
\eeaq
Inserting Eqs.(\ref{NSde2cPsi}) and (\ref{disampIII}) into this,
one can express it as
\beq
Z^{NS}_{mm'}(q',y',t')=\sum_{k=|m-m'|+1}^{m+m'-1}\chi^{NS}_{k}(q',y',t').
\eeq
This is a desired fusion rule of the neutral degenerate fields.

\subsubsection{Modular bootstrap for the (R) sector}

One can perform similar analysis for the (R) sector.
The ($\widetilde{\rm NS}$) characters are related to the (R) characters by
the following relation
\beq
\chi^{\widetilde{NS}}_{mn\omega}(q',y',t')
=\int_{-\infty}^{\infty}dP\int_{-\infty}^{\infty}d\omega'
\Psi^{R}_{mn\omega}(P,\omega')\Psi^{R\dag}_{\bf 0}(P,\omega')
\chi^{R}_{[P,\omega']}(q,y,t).
\label{Rde1PsiPsi}
\eeq
Comparing with the modular $S$-matrix element, we can find
\beq
\Psi^{R}_{mn\omega}(P,\omega')\Psi^{R\dag}_{\bf 0}(P,\omega')
=2b\sinh(2\pi mP/b)\sinh(2\pi nbP)e^{-\pi ib^2\omega\omega'}
\label{Rde1Psi}
\eeq
from which we can find
\beaq
\Psi^{R}_{mn\omega}(P,\omega')&=&
-i\sqrt{{8\over{b}}}
\left(\pi\mu\right)^{-{2iP\over{b}}}
{\Gamma\left({2iP\over{b}}\right)\Gamma\left(1+2ibP\right)
\over{\Gamma\left(ibP+{b^2\omega\over{2}}\right)
\Gamma\left(1+ibP-{b^2\omega\over{2}}\right)}}\nonumber\\
&\times&\sinh(2\pi mP/b)\sinh(2\pi nbP)
e^{-\pi ib^2\omega\omega'}.
\label{RIres}
\eeaq

It is straightforward to continue this analysis for the class-II and class-III
BCs and their mixed BCs for the (R) sector.

\section{Discussions}

In this paper we have derived conformal bootstrap equations for the $N=2$
SLFT on a half plane with appropriate boundary action and on a pseudosphere.
We have found the solutions of these functional equations which correspond
to conformal BCs.
We have also checked the consistency of these solutions by the modular bootstrap
analysis.
In particular, we have found a new class of `discrete' conformal BCs of the
$N=2$ SLFT which are parameterized by two positive integers and $U(1)$ charge.
This solutions are associated with class-I degenerate fields.
When $U(1)$ charge vanishes, it is tempting to interpret these solutions as
D0-branes in 2D fermionic black hole.
The solutions with generic integer values may describe non-BPS,
hence, unstable D0-brane.
An interesting case arises when $n=m=1$.
As we mentioned, this is different from the vacuum BC.
Our solution seems to suggest new boundary state for the 2D string theories.

Another intriguing point is the resemblance of the class-III solutions with
D0-brane solutions of the $SL(2,R)/U(1)$ coset CFT \cite{RibSch} which is dual to
the sine-Liouville theory \cite{FZZ2,KKK}.
Since the $N=2$ SLFT is dual to the fermionic $SL(2,R)/U(1)$ coset CFT \cite{GK,HoriKap},
it is natural that the two coset theories are closely related.

This relation between the coset theories means that the $N=2$ SLFT is closely
related to the sine-Liouville theory.
This can be checked by comparing the bulk reflection amplitudes.
We expect that this relationship still exists in the presence of boundary.
It is an interesting open problem to derive one-point functions based on
the conformal bootstrap of the sine-Liouville theory and
compare with the results obtained in this paper.

\section*{Acknowledgments}

We thank Al. B. and A. B. Zamolodchikov for helpful discussions.
This work was supported in part by Korea Research Foundation
2002-070-C00025.
MS is supported by the Brain Pool program from Korean Association
Science and Technology.


\end{document}